# Networked Enzymatic Logic Gates with Filtering: New Theoretical Modeling Expressions and Their Experimental Application

Vladimir Privman,[a] Oleksandr Zavalov,[a] Lenka Halámková,[c]
Fiona Moseley,[b] Jan Halámek,[c] Evgeny Katz[b]

[a]*Department of Physics, and*
[b]*Department of Chemistry and Biomolecular Science, Clarkson University, Potsdam, NY 13676*
[c]*Department of Chemistry, University at Albany, State University of New York, 1400 Washington Avenue, Albany, NY 12222*

**Abstract:** We report the first study of a network of connected enzyme-catalyzed reactions, with added chemical and enzymatic processes that incorporate the recently developed biochemical filtering steps into the functioning of this biocatalytic cascade. New theoretical expressions are derived to allow simple, few-parameter modeling of network components concatenated in such cascades, both with and without filtering. The derived expressions are tested against experimental data obtained for the realized network's responses, measured optically, to variations of its input chemicals' concentrations with and without filtering processes. We also describe how the present modeling approach captures and explains several observations and features identified in earlier studies of enzymatic processes when they were considered as potential network components for multi-step information/signal processing systems.

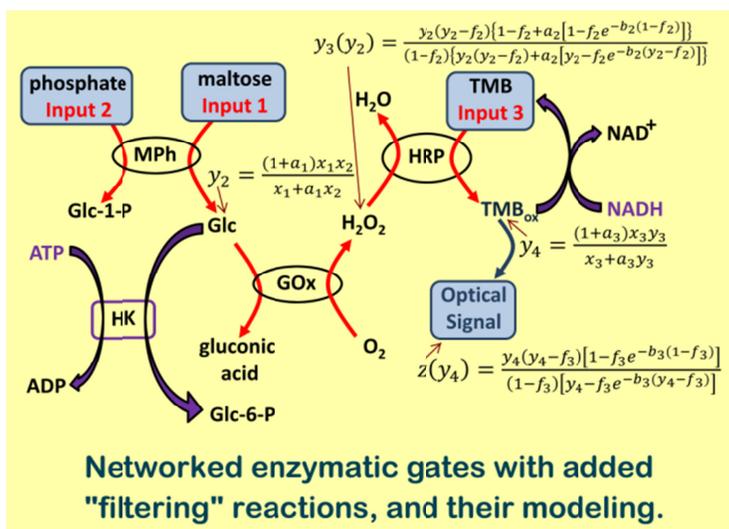

**INTRODUCTION**

Biomolecular reaction cascades offer interesting new applications as standalone systems for tailored-response[1-6] and complex signal processing,[7,8] multi-input biosensing,[9-17] and information processing[18-29] without involving electronics at each step. This offers new functionalities and applications,[30-35] including those where the output, as well as the inputs and some other process steps can be triggered, or detected as signals, by interfacing with electronics[36-44] (such as electrodes or semiconductor chips) or signal-responsive materials.[45-51] Recent results have included improvement of linear response of biosensors,[3] accomplishment of sigmoid response for certain single-input and two-input biocomputing "gates" by chemical modifications of enzymatic processes,[4-6,52-56] detection of biomarker combinations for medical diagnostics,[9-17] as well as realization of small model networks of biochemical steps for biocomputing.[18-29] Approaches to optimizing the steps (gates) and network functioning to avoid noise amplification have been developed.[22,31-33,57,58]

Biomolecular information processing ("biocomputing") systems[23,24,59,60] represent extension of recent advances in logic chemical systems[61-66] and more generally in unconventional computing.[67,68] Biocomputing systems operate with natural biomolecules: proteins/enzymes,[23,24,69,70] DNA,[27,28,30,71] RNA[72,73] and even living cells,[74,75] benefiting from their specificity and selectivity, thus allowing assembling relatively complex systems without cross-talk of their components. We have focused on enzyme-based biocomputing systems because they are particularly promising for biosensing applications[9-17] and can be easily integrated with electronic devices[36-44] and signal-responsive materials.[45-51]

Concurrently with experimental realizations, theoretical modeling ideas have been advanced[52-56] to allow few-parameter semi-quantitative description of various biochemical and added chemical processes as "gates" to be included in information/signal processing cascades. To date, there were only a few attempts[22,76] to extend and test these modeling approaches to actual networks of biochemical steps, and these did not include the latest ideas, specifically, biochemical filtering[4-6,52-56,77-79] which frequently amounts to adding simpler chemical reactions to enzyme-catalyzed processes. In this work we experimentally study a cascade of connected biochemical signal processing steps, with and without added filtering reactions, as a few-step model network. Our primary goals include theoretically deriving new fitting expressions suitable for analysis of the functioning of such networks as information/signal processing systems, elaborating the origins of parameters' dependences involved, and then testing the derived expressions against the experimental data obtained for the studied network.

**EXPERIMENTAL SECTION**

Hexokinase (HK) from *Saccharomyces cerevisiae*, EC 2.7.1.1, maltose phosphorylase (MPh) from *Enterococcus sp.*, recombinant, EC 2.4.1.8, glucose oxidase (GOx) from *Aspergillus niger*, EC 1.1.3.4, horseradish peroxidase (HRP), EC 1.11.1.7, 3,3',5,5'-tetramethylbenzidine (TMB), β-nicotinamide adenine dinucleotide (NADH) reduced dipotassium salt, adenosine 5'-triphosphate (ATP) disodium salt, maltose, sodium phosphate, glycyl-glycine (Gly-Gly) and other standard inorganic/organic reactants, such as glucose (Glc), were purchased from Sigma-Aldrich and used as supplied. Ultrapure water (18.2 MΩ·cm) from NANOpure Diamond (Barnstead) source was used in all of the experiments.

A Shimadzu UV-2450 UV-Vis spectrophotometer with a TCC-240A temperature-controlled holder and 1 mL poly(methyl methacrylate) (PMMA) cuvettes was used for all optical measurements. All experiments and optical measurements were performed in 0.05 M Gly-Gly buffer, pH = 7.3, at 40.0 ± 0.2°C, also used as the reference background solution. Scheme 1 shows the sequence of biocatalytic processes involved in the enzymatic cascade. The system is (bio)catalyzed by the three enzymes, and one of the added filter processes involves an additional enzyme. The main, non-filter, steps are as follows (Scheme 1). MPh catalyzes the conversion of maltose and inorganic phosphate into β-D-glucose-1-phosphate (Glc-1-P) and glucose. This is followed by glucose oxidation catalyzed by GOx in the presence of oxygen, to form gluconic acid and hydrogen peroxide. Hydrogen peroxide reacts with TMB in the presence of HRP to form a blue colored oxidized product, $TMB_{ox}$, the concentration of which was measured at 655 nm.

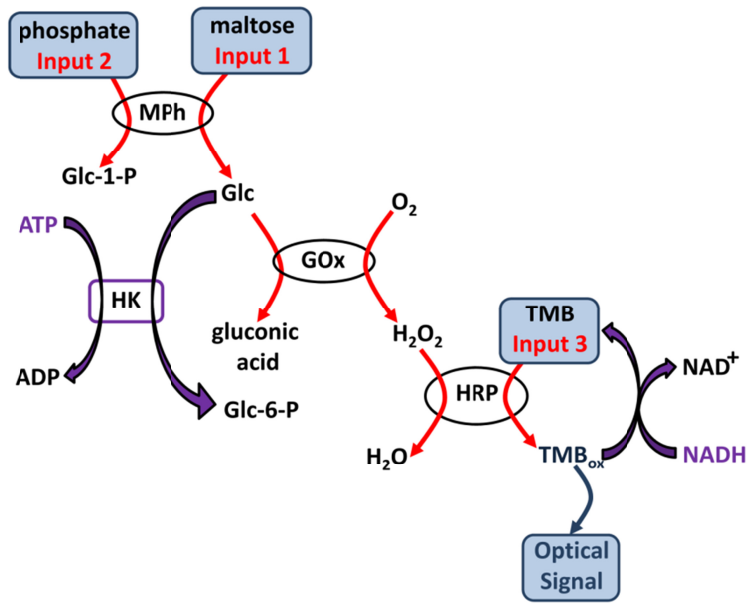

**Scheme 1.** The biocatalytic cascade with three variable inputs (maltose, phosphate, TMB), and two optional added "filtering" processes, one biocatalyzed by hexokinase, and another involving the "recycling" of the output chemical (TMB$_{ox}$) by NADH. Abbreviations for various chemicals are defined in the text.

To map out the response of the biocatalytic cascade to the initial concentrations of substrates selected as inputs (Scheme 1), maltose, phosphate and TMB were varied starting at 0, as the reference logic-**0** values, and increasing up to conveniently selected reference logic-**1** values. For maltose, phosphate and TMB, these were 9.0 mM, 11.0 mM, and 0.8 mM, respectively. In addition to the input substrates, the non-input "gate machinery" reactants were dissolved in the solution at the following initial concentration: MPh (2 U/mL), GOx (2 U/mL), HRP (0.2 U/mL). The filter-process chemicals, when added, had the initial concentrations: HK (2 U/mL), ATP (1.25 mM), when the HK-filter was activated, and NADH (0.1 mM) when the NADH-filter was activated, separately or both together; see Scheme 1. Byproducts produced in the HK-filter process are adenosine diphosphate (ADP) and α-D-glucose-6-phosphate (Glc-6-P), whereas the NADH-filter process produces β-nicotinamide adenine dinucleotide (NAD$^+$).

The experiments were performed to study the system's response to the variations of each of the three input substrate concentrations with the other two substrates initially at their maximal concentration, without added filter processes, and then repeated with the added HK-filter process, separately with the added NADH-filter process, and also with both filter processes added. This yielded 12 data sets for the output recorded at the "gate time" set at 420 sec, as the absorbance, Abs, at the absorption peak of the oxidized TMB at 655 nm.

**THE SYSTEM'S FUNCTIONING AS A MODEL NETWORK**

In this section we outline the functioning of our system as an information/signal processing network. We use it as a model system to explore ideas of parameterizing and optimizing small-network functioning. Furthermore, the present system is of interest because it consists of steps similar to those which have also been incorporated in enzymatic cascades devised for biosensor application involving detection of maltose or starch.[80-84] The (bio)chemical processes in our system are shown in Scheme 1. The first step functions as an AND logic gate with two variable inputs: maltose, which we select as logic Input 1, and phosphate, selected as Input 2. Scheme 2 shows this and other steps in the system interpreted as binary "logic gates" and non-binary (analog) "filtering" functions addressed in the next paragraph. The output, Glc, of the first AND gate is an input for the second enzyme, GOx, the action of which biocatalyzes the production of $H_2O_2$. This can be considered an identity binary "gate", denoted by I. The produced $H_2O_2$ in turn is an input for the third enzyme, HRP, which then uses TMB, selected as logic Input 3, to yield the output chemical product, $TMB_{ox}$, as another AND gate function. The final output signal, Abs, is measured optically as described in the preceding section. In our analysis in the next section, this optical measurement of the chemical concentration of $TMB_{ox}$ can be viewed as another I-gate step in the network.

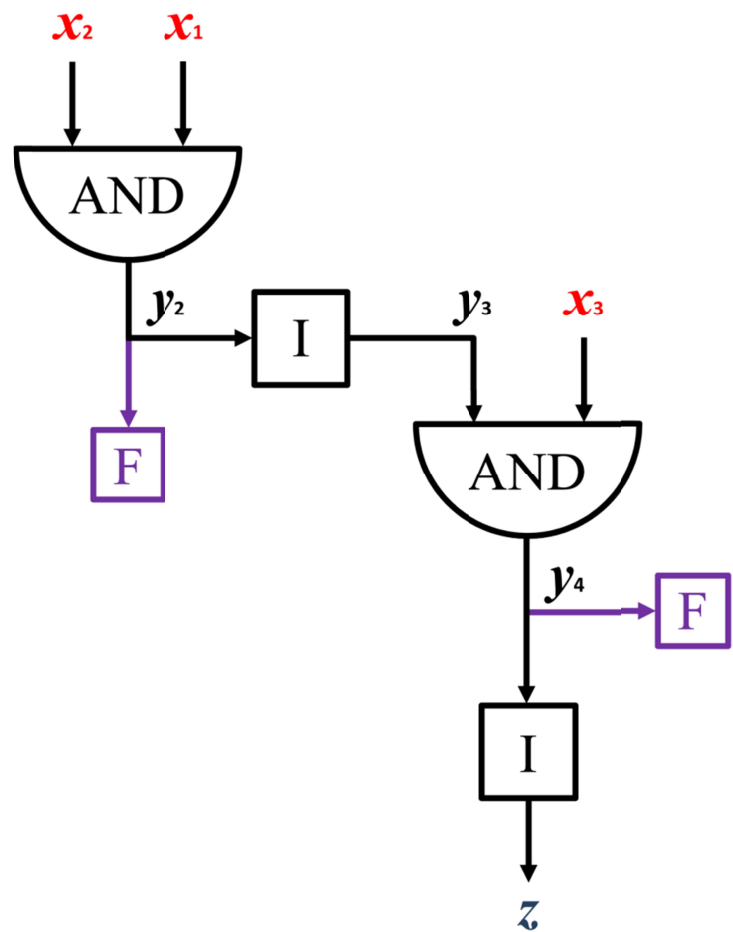

**Scheme 2.** Interpretation of the (bio)chemical processes shown in Scheme 1, as a "network" of binary AND and I gates, as well as non-binary "filtering" functions.

Two different "filtering" processes, marked by F, can be optionally added, singly or together, one biocatalyzed by HK, the other involving the "recycling" of the output chemical by NADH. These are shown in our "network" interpretation, Scheme 2. We will address each of the network steps separately later. We point out, however, that one of the interesting features of the present network is that it can be made more complicated in future studies, for example, as illustrated in Scheme 3. The middle processing step can be made into an AND-gate function by controlling the supply of oxygen. In fact, the functioning of GOx as an AND gate, with two inputs being Glc and oxygen was already utilized in another study.[85] Another "filtering" process can be added for one of the first-step inputs, phosphate. As detailed in the insets in Scheme 3, this can be accomplished by adding an enzymatic process biocatalyzed, for instance, by glycogen phosphorylase (GPh), EC 2.4.1.1,[86] or another enzyme that competes for phosphate as an input, or by adding a chemical step, where phosphate ions can be precipitated, for example, by adding magnesium cations, yielding insoluble magnesium phosphate, predominantly $Mg_3(PO_4)_2$ at ca. pH 8.[87,88] Both of these processes involving phosphate are not expected to cross-react with other steps in the system.

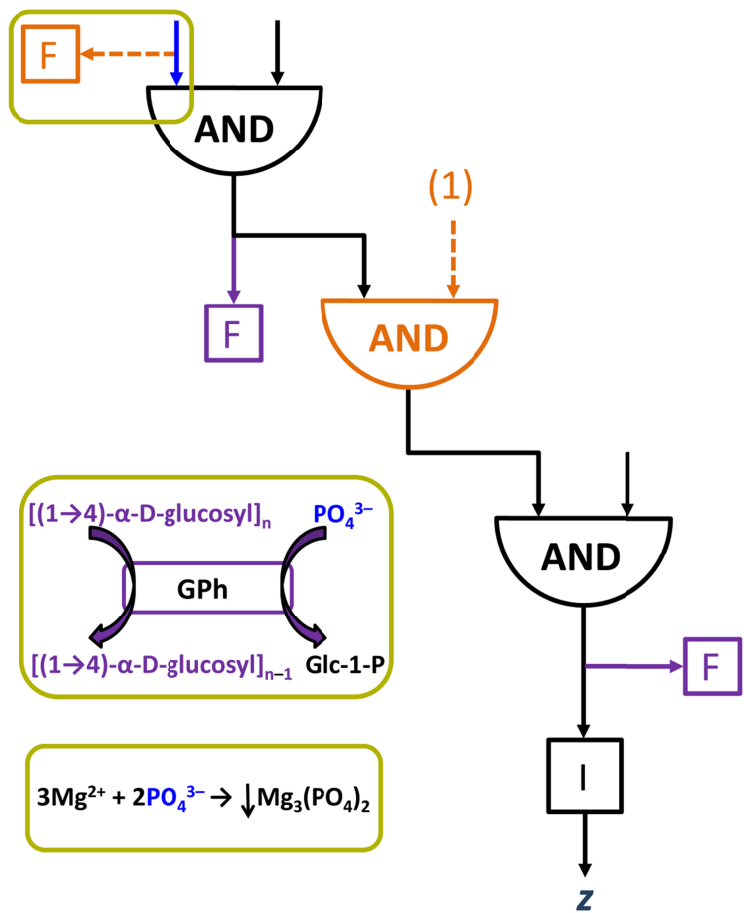

**Scheme 3.** An alternative more complicated network that can be realized in possible future studies, as described in the text. The middle processing step can be made an AND gate, with oxygen varied as the second input (here kept fixed, marked "1"). Another optional filtering step can be added for one of the first-step inputs, phosphate (shown in the scheme for simplicity as $PO_4^{3-}$, but could appear in protonated forms depending on pH values). The insets show possible enzymatic or chemical realizations of this added filter (where GPh and Glc-1-P denote glycogen phosphorylase and glucose-1-phosphate, respectively).

The present system involves enzymes some of which have complicated mechanisms of action, as commented on later. For example, for MPh the specifics of the mechanism are not well studied, and the order of intake of the two substrates is not unique.[89,90] In the next section, we describe the motivation for and the details of a simplified modeling approach suitable for evaluation of such systems as logic-gate networks. Here we comment that the designation of the "logic inputs," such as maltose and phosphate as Inputs 1 and 2, for information processing is made based on the desired application and does not imply that this is the actual kinetic order of their intake. All three inputs are varied from some application-determined logic-**0** values, here taken as the initial concentrations 0 for convenience, to logic-**1** values. The latter were selected as experimentally convenient values for our present model study, but generally will also be application-determined. For analysis of the system's functioning as a logic network, we then define scaled variables in the range from **0** to **1**, here, for example,

$$x_1 = [\text{maltose}](t = 0)/[\text{maltose}]_{\max}, \tag{1}$$

where $t$ denotes the time, and $[\text{maltose}]_{\max}$ is the maximum (logic-**1**) initial concentration selected for Input 1, here 9.0 mM. Variables $x_2$ and $x_3$ are defined similarly. For the output signal, we define

$$z = \text{Abs}_{\mathbf{000}}(t = t_g)/\text{Abs}_{\mathbf{111}}(t = t_g), \tag{2}$$

where the absorbance of the generated TMB$_{\text{ox}}$ is measured at the gate time, $t_g$, with the reference logic values set by the system functioning: logic-**0** at zero inputs, **000**, and also for inputs **001**, **101**, etc., totaling seven combinations with at least one zero, and logic-**1** at inputs **111**. The definition of the logic variables $y_{2,3,4}$ for the intermediate products are also similar, but for $y_{2,3}$, in particular, they require additional discussion because of time dependence. We will address this in the next section.

Our goal in modeling networks like the one considered here for purposes of evaluating their utility as information processing systems, is to devise an approximate description which suffices to parameterize the "response shape," here, the function $z(x_1, x_2, x_3)$. We seek a description with as few parameters as possible, which can be approximate as long as it offers information on selected features of the response. For binary "gates" based information processing, we seek to decrease noise amplification or, better, achieve noise suppression in the vicinity of the logic-point values of the inputs. The noise-spread transmission factor, assuming approximately equal spread of noise in all the inputs when normalized per their "logic" ranges, can in most cases be estimated by the absolute value of the gradient, here

$$|\vec{\nabla}z| = \sqrt{\left(\frac{\partial z}{\partial x_1}\right)^2 + \left(\frac{\partial z}{\partial x_2}\right)^2 + \left(\frac{\partial z}{\partial x_3}\right)^2} . \qquad (3)$$

Ideally, the largest value of this quantity when calculated near all the logic points should be less than 1. At a single-gate level, the added chemical "filtering" steps can facilitate this.[52,54,55] For other applications, such as sensor design,[3] one might instead seek other adjustments of the response function properties, such as achieving linear response of the output with respect to varying one of several inputs.

The function $z(x_1, x_2, x_3; ...)$ parametrically depends on various quantities (denoted by ...) which are not the scaled inputs $x_{1,2,3}$, but are other chemical or physical properties that can to some degree be adjusted by chemical or physical means to modify the system's response. These include initial (bio)chemical concentrations of reactants which are not the inputs or measured as the output, and process rates (which depend on the chemical and physical conditions of the system). An advantage of considering the "logic" scaled variables for optimization specifically for enzymatic systems is that less fitting parameters are involved, as will be explained in the following sections.

However, not all optimization tasks can be carried out in the "logic" language. The most obvious counterexample involves avoiding the loss of the overall signal intensity, here, the spread between $\text{Abs}_{111}(t = t_g)$ and $\text{Abs}_{000}(t = t_g)$ in the notation of Equation (2), which can result from the added "filtering" processes. Furthermore, the mere possibility of the optimization by "tweaking" the network to change the "analog" information processing responses of it as a whole or its constituent "gates," is usually limited to networks which are not too large. For large enough networks "digital" optimization will ultimately be required,[91] involving the redesign of the network with trade-offs involving redundancy, in order to avoid noise buildup.

**THEORETICAL SECTION**

*Phenomenological modeling of network elements*

Biocatalytic processes considered as "gates" within multistep signal processing cascades can be modeled at various levels. Individually, enzymatic reactions themselves involve several steps and can be rather complicated and have various pathways of functioning, some of which are not fully understood and can actually vary depending on the source of the enzyme and other parameters. In our case, the mechanism of action of MPh is complicated and not well studied,[80-82,92] whereas GOx has a relatively well understood and straightforward mechanism.[93] HRP has a generally-known, but rather complicated mechanism of action,[94] while HK has a non-unique order of intake of its substrates.[95] In our context of signal processing networks, it would be impractical to attempt to use the full-complexity kinetic modeling involving multiple rate-constant parameters for each of the involved enzymes. The available data are not detailed enough for an accurate kinetic description. Furthermore, such accuracy is not required because our goal is to describe the function $z(x_1, x_2, x_3)$ semi-quantitatively,[55] in order to evaluate and if needed adjust its behavior in the vicinity of the logic values of the inputs to improve the network's noise handling[22,31-33,57,58] properties. This can be accomplished by using an approximate, few-parameter fitting for each step of the signal processing,[54,55] or, as the network becomes larger, by adopting a more engineering approach of entirely phenomenological fitting expressions[22,56] that reproduce the generally expected features of the function $z(x_1, x_2, x_3)$. However, ideally a hybrid approach should be favored whereby the phenomenological fitting expressions are derived[56] from simplified kinetic considerations for each sub-process in the network. This allows making a connection between the phenomenological fit parameters and physical/chemical properties (such as rates or concentrations), thus enabling better control of the network's functioning by adjusting these parameters. Here we use this approach, relying on earlier works[22,56] and also deriving and systematizing new expressions, for the first time for a biochemical network of the present complexity.

*Networked AND gates without filtering*

Here we use ideas developed[56] in the context of an "identity gate" (signal transduction) of using a Michaelis-Menten (MM) like approximate description[96-98] of enzymatic reactions and additional approximations suitable for "logic-gate" modeling. We derive a new, rather surprising result for parameterizing two-input (two-substrate) AND gates of the type used in our network, which have generally been the most popular standalone biocatalytic logic gates realized with enzymes.[22,52,55-57,99-102] We use a simplified MM kinetic scheme representing the main pathway for the action of the considered enzyme,

$$S + E \xrightarrow{k_S} C ,\qquad(4)$$

$$U + C \xrightarrow{k_U} E + P ,\qquad(5)$$

where the enzyme, $E$, first binds the substrate, $S$, to form a complex, $C$, which later reacts with the other substrate, $U$, to yield the product, $P$. As common in considering logic-gate functioning,[55,79] we ignored a possible back-reaction,[103,104] with rate constant $k_{-S}$, in Equation (4), to decrease the number of adjustable parameters, and also because in such situations large quantities of the substrates are typically used (at least for logic-**1** values) to "drive" the process to yield large output range. We will revisit this approximation later.

We note that enzymatic reactions typically function in an approximate steady state for extended time intervals.[96-98] This is not always the case, and in fact, a very fast reaction regime of saturation was shown to allow avoiding noise amplification in some situations.[58,76] However, this requires special parameter optimization. Since our network parameters were experimentally conveniently but otherwise randomly selected, we assume generic behavior for its sub-processes.

Specifically, for a two-input process of the type modeled by Equations (4-5), in the steady state the fraction of the enzyme in the complex is approximately constant, and we can assume that

$$\frac{dC}{dt} = k_S SE - k_U UC = k_S SE - k_U U(E_0 - E) \approx 0, \qquad (6)$$

where the subscripts 0 will denote values at time $t = 0$. Therefore, in the steady state we expect

$$E \approx \frac{E_0 k_U U}{k_S S + k_U U}, \qquad (7)$$

and thus

$$\frac{dP}{dt} = k_U UC \approx \frac{E_0 k_S S k_U U}{k_S S + k_U U}. \qquad (8)$$

Since in signal processing applications the reaction is usually driven by the availability of substrates, we can ignore their depletion and write the following approximate expression for the rate of the product generation and for its total quantity produced at $t = t_g$,

$$\frac{dP}{dt} \approx \frac{E_0 k_S S_0 k_U U_0}{k_S S_0 + k_U U_0}, \qquad (9)$$

$$P(t_g) \approx \frac{E_0 k_S S_0 k_U U_0 t_g}{k_S S_0 + k_U U_0}. \qquad (10)$$

While several assumptions were made to yield this result, we point out that the resulting expressions are typical of the steady-state-type MM approximations, and were also used successfully[56] to fit data for a single-input "identity gate function" case. Here we consider a two-input AND gate, and therefore the logic-variable description will involve the function $z(x, y)$, with the variables defined according to

$$z = P(t_g)/P(t_g)_{max}, \quad x = S_0/S_{0,max}, \quad y = U_0/U_{0,max}, \qquad (11)$$

where the subscript *max* refers to the largest (logic-**1**) values. These satisfy the same relation, Equation (10), and therefore substantial parameter cancellations occur as we divide the general Equation (10) by its logic-**1** counterpart, to yield our final expression

$$z(x,y) = \frac{(1+a)xy}{x+ay}, \quad (12)$$

with

$$a = \frac{k_U U_{0,max}}{k_S S_{0,max}}. \quad (13)$$

This is a rather interesting result, because it suggests that the logic-gate functioning as an AND function, for enzymatic systems in the considered regime can be approximately parameterized with just a *single adjustable parameter*, denoted *a* in Equations (12-13). In fact, this conclusion captures many empirical observations reported earlier for such "non-filtered" gates, when more sophisticated fitting schemes involving kinetic descriptions[54,55] or two-parameter[22] entirely phenomenological expressions were used. Specifically, it was found[22] that it is difficult to affect the logic-function properties by changing the amount of enzyme or the gate time, which is now explicit in the developed approximations because these quantities ($E_0$ and $t_g$) entirely cancelled out of the expression for *a* in Equation (13). On the other hand, the logic-**1** values of the two inputs (which are set by the environment in which the gate operates) do affect the shape of the response surface. This is shown in Scheme 4, which illustrates possible surfaces described by Equation (12). We note that interchanging the labeling of the inputs, $x \leftrightarrow y$, corresponds to replacing $a \leftrightarrow 1/a$, so that the $a = 1$ case is the most symmetrical. All such gates are convex and amplify noise, with the noise transmission factor, i.e., the maximal slope of $z(x,y)$ among the four logic points, equal $1 + \max(a, a^{-1})$ in the context of our parameterization. It assumes its smallest value, 2 in the symmetrical case, i.e., 200% noise amplification. This is typical[57] of non-optimized standalone enzymatic AND gates of this sort. For asymmetric cases, cf. Scheme 4, the noise amplification factor, $1 + \max(a, a^{-1})$, can far exceed 2.

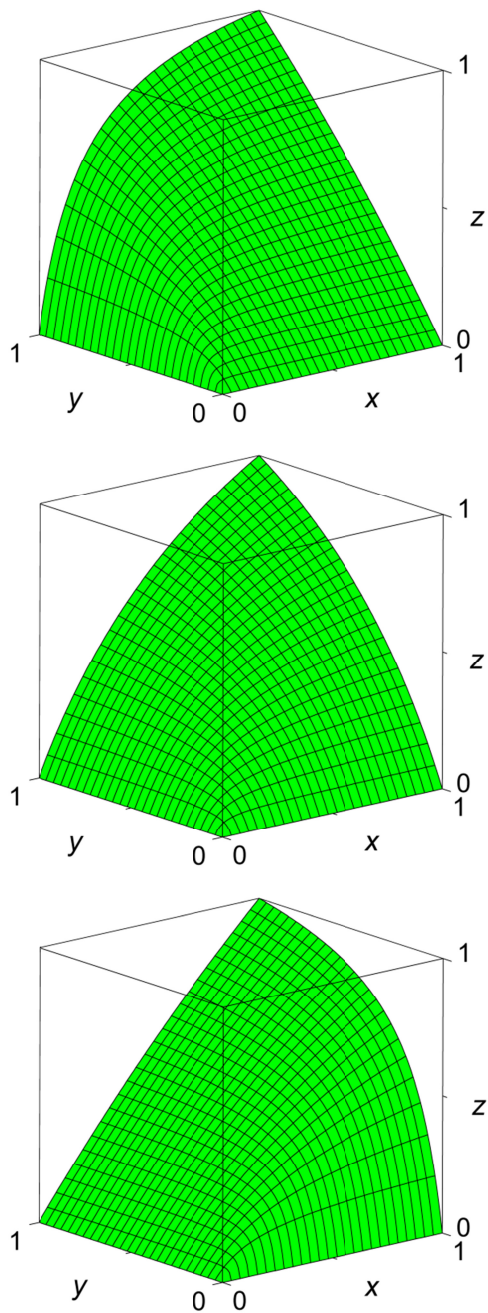

**Scheme 4.** The function shown in Equation (12) for $a = 1/5$ (top panel), 1 (middle panel), and 5 (bottom panel).

We comment that, with $y = 1$, i.e., with only one varied input, Equation (12) reduces to the earlier studied[56] parameterization of the single-input "identity gate." We now combine the approximate kinetic expressions of the type shown in Equation (12) for the steps in our system, and we then discuss possible limitations of such approach. In the notation of Scheme 2, for each step, except for the last "identity gate" which is assumed approximately linear ($z = y_4$), a distinct parameter $a$ is introduced,

$$y_2 = \frac{(1+a_1)x_1x_2}{x_1+a_1x_2}, \tag{14}$$

$$y_3 = \frac{(1+a_2)y_2}{1+a_2y_2}, \tag{15}$$

$$z = y_4, \quad y_4 = \frac{(1+a_3)x_3y_3}{x_3+a_3y_3}. \tag{16}$$

Concatenating these relations to describe the function $z(x_1, x_2, x_3)$ can be questioned, because the successive steps (gates) feed one another, and therefore intermediate products are time-dependent. However, considering that within the present assumptions the product generation in each step is irreversible, cf. Equation (5), and all the concentrations "driven" by each gate's inputs are linear in the gate-time, Equation (10), the concatenation can be a reasonable approximation,

$$\begin{aligned} z = (1 + a_1)(1 + a_2)(1 + a_3)x_1x_2x_3 / \\ [x_1x_3 + a_1a_2a_3x_1x_2 + a_1a_2x_1x_2x_3 + \\ a_1a_3x_1x_2 + a_2a_3x_1x_2 + a_2x_1x_2x_3 + \\ a_1x_2x_3 + a_3x_1x_2]. \end{aligned} \tag{17}$$

We will use variants of this expression for data fitting in the next section, as well as offer additional discussion. First, however, in the remainder of this section we consider the added filtering processes.

*Incorporation of filtering steps in networks*

Phenomenological modeling of added filtering processes by approaches of the type considered here is rather recent,[56] and thus far has only been reported for a single-input "identity gate," with the added "intensity filtering" process deactivating part of the input by utilizing a competing chemical reaction.[4-6,105] This added process then converts the convex response to sigmoid. Other phenomenological descriptions are possible,[5,6,22,32,106-108] notably, the Hill-function fitting,[106-108] which, however, is more suitable in situations of sigmoid response being caused by cooperativity, for instance, when enzyme allostericity or similar effects are involved.[1,2,109,110]

Our first "filter" process competes for the input (Glc) of the enzyme GOx, see Scheme 1, and therefore can be regarded as functioning as described above. Indeed, the concentration of oxygen is not a varied input, and therefore its concentration can be lumped with the rate constant $k_U$ into a single fixed rate-constant-type parameter combination $\overline{k_U} = k_U U_{0,max}$ that enters phenomenological expressions such as Equation (13). The added filtering process biocatalyzed by HK, Scheme 1, then competes for a fraction, $F_0$, of the input Glc, up to $S_{0,max}$. This depletion due to the diversion of part of the input is phenomenologically modeled in a simplified fashion by adding the process

$$S + F \xrightarrow{k_F} \dots , \qquad (18)$$

where $F$ is initially set to $F_0$. The parameters $F_0$ and $k_F$ are phenomenological because this is a very approximate description rather than a realistic kinetic modeling of the added HK step (Scheme 1). For the considered case, $F_0$ can be approximately adjusted by varying the initial concentration of ATP, whereas the overall process rate constant, lumped in $k_F$, can be varied by changing the amount of HK. This crude approximation aims at obtaining a simple fitting expression without attention to the details of the actual kinetics. The process Equation (18) alone, by the gate time $t_g$, would deplete[56] the availability of the substrate $S$ according to

$$S(t_g) = (S_0 - F_0)S_0/[S_0 - F_0 e^{-k_F(S_0-F_0)t_g}] \ . \tag{19}$$

We then use[56] this expression as accounting for the reduced intensity, to replace $S_0$ in Equation (10), with $U_0$ set to $U_{0,max}$, to write

$$P(t_g) \approx \frac{k_S k_U U_{0,max} E_0 t_g}{k_S + \frac{k_U U_{0,max}}{(S_0-F_0)S_0}[S_0 - F_0 e^{-k_F(S_0-F_0)t_g}]} \ . \tag{20}$$

In terms of the scaled variables for this step, see Scheme 2, and its earlier introduced parameter $a = a_2$, we can then obtain the expression to replace Equation (15),

$$y_3(y_2) = \frac{y_2(y_2-f_2)\{1-f_2+a_2[1-f_2 e^{-b_2(1-f_2)}]\}}{(1-f_2)\{y_2(y_2-f_2)+a_2[y_2-f_2 e^{-b_2(y_2-f_2)}]\}} \ , \tag{21}$$

Except for relabeling the scaled variables and adding index 2 to the fitting constants to designate the gate, this is essentially the same expression as derived in earlier work,[56] with the general relations for the new fitting parameters (without the index 2),

$$f \equiv F_0/S_{0,max} \ , \quad b \equiv k_F S_{0,max} t_g \ . \tag{22}$$

Note that we expect the values of phenomenological parameters defined in this section to generally satisfy $a > 0$, $0 \leq f < 1$, $b \geq 0$, for each step that they are introduced for. In addition to the fact that for individual gates, added filtering processes frequently improve noise-transmission properties by making their response sigmoid in one or both inputs, these processes are also useful in the general context of modifying network functioning. Indeed, they are easier to utilize for control and modification of the network response, because the parameter $f$ can be adjusted by varying the amount of the supplied "filtering" chemical (here, ATP), whereas the parameter $b$ can be changed not only by varying the process rate (here, by amount of HK) but also directly by selecting the gate time, $t_g$, cf. Equation (22). Plots of functions such as Equation (21) for representative parameter values were given in earlier work.[56]

In the preceding discussion, as well as in earlier work,[56] we avoided modeling of the added filter process for two-input AND gates, because the situation in this case is more complicated and it is not known whether the approach just described can be extended to yield straightforward, few-parameter analytical expressions, such as Equation (21). Such simple, analytical expressions, which are obtained supplemented with some kinetic interpretation of the involved parameters, see our Equations (13, 22), are particularly convenient if we seek description of multi-step networks for which a more detailed, realistic kinetic modeling is not a viable alternative due to its complexity. In the present system, the third (HRP) step of the processing, see Scheme 1, with the added chemical filter of the output involving "recycling" one of the input substrates by the added NADH, is such an output-filtered two-input AND gate. We bypass the afore-described difficulty of modeling it directly, by considering it as a part of the network in which, as shown in Scheme 2, we in advance somewhat artificially singled out the chemical-to-optical signal conversion as an additional *single*-input "identity gate." We consider the added filter process as competing for the input, $TMB_{ox}$, of this step, which was earlier regarded as approximately linear. We note that linear response is obtained as the limit of large $a$ in our phenomenological modeling of single-input identify functions, cf. Equation (15) for a different step. Therefore, we adopt the $a \to \infty$ limiting form of the expressions with filtering, such as Equation (21), instead of the final-step linear function, see Equation (16), i.e., we take

$$z(y_4) = \frac{y_4(y_4 - f_3)[1 - f_3 e^{-b_3(1-f_3)}]}{(1-f_3)[y_4 - f_3 e^{-b_3(y_4-f_3)}]}, \qquad (23)$$

but the relation for $y_4(y_3)$ in Equation (16) remains unchanged. Here subscript 3 designates the two added fitting parameters, $f_3$ and $b_3$, of to the filtering process involving NADH reacting with the output of the third gate in the original cascade, consistent with the notation for $a_3$ for that gate. The parameter $f_3$ can be approximately adjusted by varying the NADH concentration, whereas $b_3$, related to the rate constant, can be changed by adjusting the gate time.

Various relations derived in this section can be concatenated to write down expressions which replace the "no filters" Equation (17) with appropriate formulas for the cases of one or both of the filtering processes shown in Scheme 1 added. These analytical expressions are too cumbersome to display explicitly. However, we point out that the concatenation can be done in a computer, and the whole network description is easily programmed for data fitting, the results of which are described and discussed in the next section.

**RESULTS AND DISCUSSION**

Our main goal in this work has been to establish that the proposed parameterizations of the individual steps, when concatenated, can offer a reasonable description of the network's functioning. These parameterizations should be used with some care, as addressed later. Furthermore, even if taken literally they involve 7 fitting parameters: $a_{1,2,3}$, $f_{2,3}$ and $b_{2,3}$. Fitting these all at once is impractical. However, we will demonstrate that by probing network response to individual inputs we can determine parameter values one or two at a time. Let us first consider the network without filtering. We set two inputs at a time at their logic-**1** values, which were 9.0 mM, 11.0 mM, and 0.8 mM, for maltose (Input 1), phosphate (Input 2) and TMB (Input 3), respectively. We then varied the remaining input from 0 to its logic-**1** concentration and measured the network's output. The results, scaled to the logic variables, are shown in Figure 1. The logic-**1** value for the output depends on the gate functioning, which in itself is slightly noisy from one realization to another, and was, in this case, averaged over slightly fluctuating (within few percent) experimental values, $\text{Abs}_{\mathbf{111}}(t_g) = 1.79$.

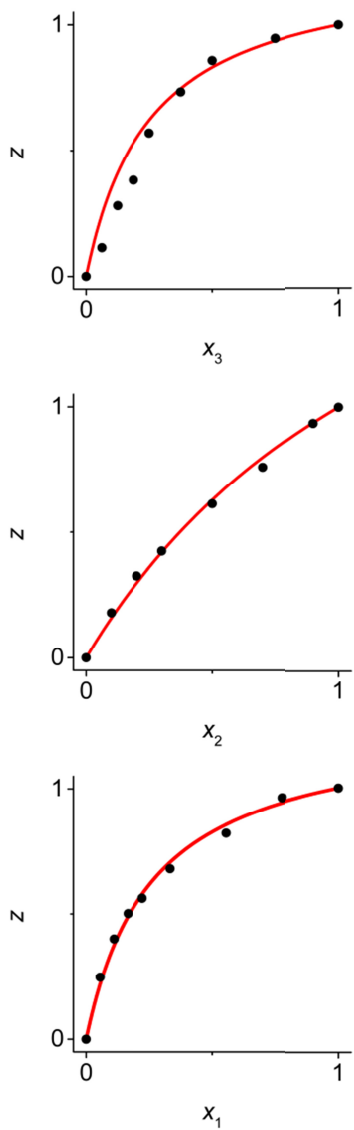

**Figure 1.** Experimental data (points) and single-parameter fitted curves, see Equations (24-27), for the variation of the absorbance for the network without any filter processes added, as a function of the three externally controlled inputs identified in Scheme 2. All the varied concentrations (of the inputs) and measured signal (the absorbance) were expressed in terms of the dimensionless logic-range variables, such as those defined in Equations (1-2).

The data for varying Input 3 (Figure 1) can be fitted by substituting $x_{1,2} = 1$ in Equation (17), which yields an expression that only depends on a single parameter, $a_3$,

$$z(x_3) = \frac{(1+a_3)x_3}{x_3+a_3}. \tag{24}$$

Least-squares data fit then gives $a_3 = 0.26$. We now consider the variation of Input 2 (see Figure 1), for which setting $x_{1,3} = 1$ in Equation (17) gives a result that involves on a single new combination of parameters,

$$z(x_2) = \frac{(1+a_3)(1+A_{12})x_2}{1+[(1+a_3)A_{12}+a_3]x_2}, \tag{25}$$

$$A_{12} \equiv a_1 + a_2 + a_1 a_2. \tag{26}$$

By using the known value of $a_3$, we fitted the data to get $A_{12} = 0.47$, see Figure 1. We next put $x_{2,3} = 1$ in Equation (17), to get the expression

$$z(x_1) = \frac{(1+a_3)(1+A_{12})x_1}{a_1+[(1+a_3)(1+A_{12})-a_1]x_1}. \tag{27}$$

Again, with $a_3$ and $A_{12}$ known, only a single new parameter is involved, fitted to give $a_1 = 0.31$. Finally, $a_2$ is calculated from Equation (26), $a_2 = 0.12$. We conclude that our phenomenological approach offers a reasonable fitting of the data without filtering. We will now use the determined parameter values for $a_{1,2,3}$ in data fitting with filter(s) added.

Let us first only add the HK-catalyzed filter; see Schemes 1 and 2. Again, we probe the network's response to each of the three inputs separately, with the fixed inputs at their logic-**1** values. The results are shown in Figure 2. The average value of the logic-**1** output in this case was $\text{Abs}_{\mathbf{111}}(t_g) = 1.65$, with the level of noise again only within a few percent. It was important to adjust the "intensity" of this filtering process at a moderate enough level such that the overall intensity of the signals in the network is not significantly decreased. Otherwise, we could not use the "unfiltered" network parameters, $a_{2,3}$, estimated earlier, for the "filtered" data. Indeed, Equation (13) suggests that decrease in the availability of certain substrates as inputs for the intermediate steps of the signal processing can affect the values of these parameters. Here the loss of intensity was insignificant (from average maximal absorbance 1.79 to 1.65) on the scale of the overall degree of noisiness of the experimental data.

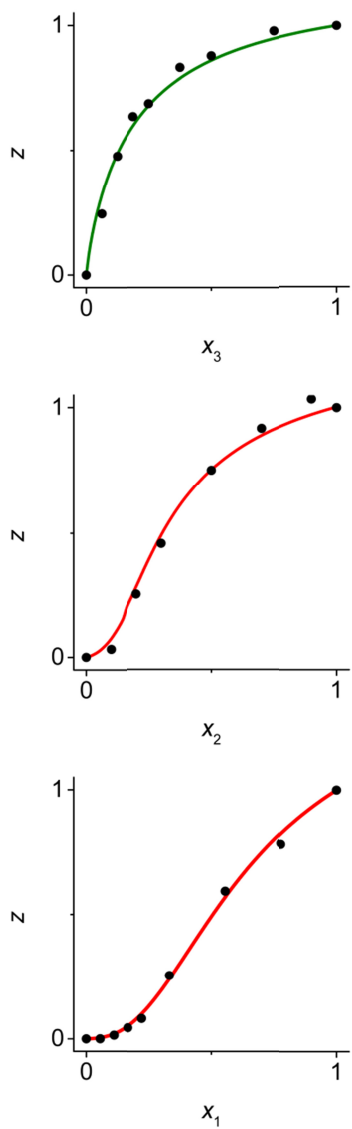

**Figure 2.** Experimental data (points) and theoretical curves for the variation of the absorbance as a function of different inputs, with all the quantities normalized to their logic ranges, similar to Figure 1, but with the HK-catalyzed filter process added, see Scheme 2. Note that the theoretical curve in the top panel is not fitted but rather calculated by using a parameter estimate obtained earlier; see text for details.

Therefore, we can use the earlier estimated parameters, $a_{1,2,3}$, for the HK-filtered system, results for which are reported in Figure 2. The logic-variable response to $x_{1,2}$ involves also the dependence on the parameters $f_2$ and $b_2$. The explicit function is too complicated to be displayed. For computer evaluation, it was programmed by concatenating separate processing-step expressions derived in the preceding section. We used a simultaneous least-squares fit of both data sets shown in the two bottom panels of Figure 2, to estimate $f_2 = 0.42$, $b_2 = 7.9$. The top panel shows data which, in terms of the logic variables, should still be described by Equation (24), provided the output and intermediate signal intensities were not much reduced, as explained in the preceding paragraph, so that we can use the earlier estimated value of $a_3$. The curve shown in the figure was drawn without any data fitting, by using Equation (24).

We now consider the addition of only the NADH filter; see Schemes 1 and 2. The results are presented in Figure 3. The average logic-**1** output value in this case was $\text{Abs}_{\mathbf{111}}(t_g) = 1.14$, indicating a notable reduction as compared to the unfiltered case, and the spread of the three values was also larger, about 13%, illustrating that enzymatic networks of this degree of complexity can in some regimes be rather noisy. Here, however, the fact that this filtering process decreases the output intensity does not affect our model in terms of the logic variables, because the added chemical reaction occurs at the input of the last, "identity gate" step, see Scheme 2, which was already assumed approximately linear. Decrease in the $\text{TMB}_{\text{ox}}$ concentration can only make it conversion to absorbance more linear. However, more generally in the context of networked biochemical steps, loss of the overall signal intensity can be an undesirable tradeoff of the added control and sometimes sigmoid response options offered by added filtering processes, because this can make the ever-present noise more significant of the scale of the useful signal variation. In this case all three curves depend on the parameters $a_{1,2,3}$ and $f_3$, $b_3$. We used the response curves to the variation of $x_{1,2}$ (two bottom panels) to fit these two parameters: $f_3 = 0.31$, $b_3 = 42$. The top-panel curve in the figure was not fitted. It was drawn using the already estimated parameters; it actually only involves $a_3, f_3, b_3$.

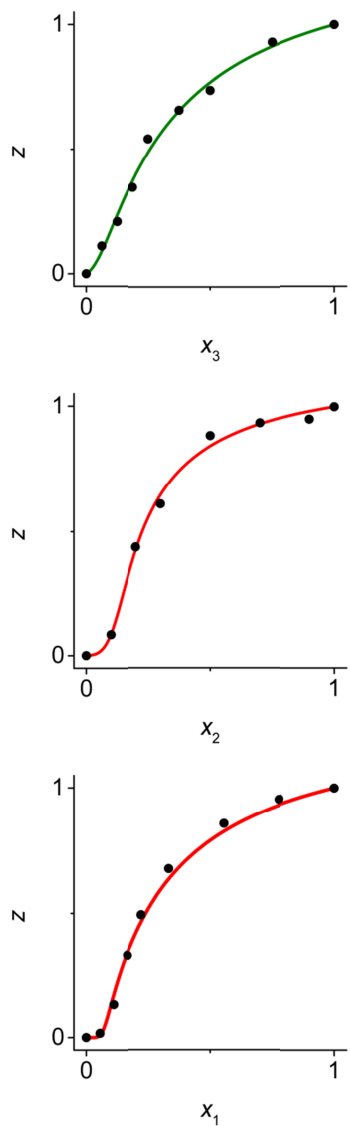

**Figure 3.** Experimental data (points) and theoretical curves for the variation of the absorbance as a function of different inputs, with all the quantities normalized to their logic ranges, similar to Figures 1, but with the NADH filter process added, see Scheme 2. Note that the theoretical curve in the top panel is not fitted but rather calculated by using parameter estimates obtained earlier; see text for details.

Finally, let us consider data fitting with both filter processes active. These results are shown in Figure 4. In this case the average maximum output intensity was $\text{Abs}_{\mathbf{111}}(t_g) = 0.47$, practically the same value for all three input variations. While the fact the intensity dropped this much is not of concern for the logic-variable analysis, it implies that the actual noise in the data will be more significant on the *relative* scale. This is clearly seen in the figures, with the data in all its three panels being noticeably noisier that in the earlier-considered cases. We note that in this case no new parameters are involved, and therefore, all the theoretical curves shown in Figure 4 were calculated by using the earlier estimated values. The curves of the dependence on $x_{1,2}$ (two bottom panels) require all the gate-function and filter parameters for their evaluation. The explicit formulas are too cumbersome to display, but, as mentioned earlier, they can be straightforwardly programmed for computer evaluation. The $x_3$ dependence here (the top panel) is, in terms of the logic variables, identical to that in the top panel of Figure 3, i.e., the theoretical curves are the same, and it only involves $a_3, f_3, b_3$.

Considering the relatively noisy data in this case, our semi-quantitative fit in terms of the logic variables works quite well. Possible improvements, especially to address the notable fit vs. data mismatch in the middle panel, can perhaps be achieved by utilizing an additional parameter offered by considering the possible reversibility of the first step in the MM description for the first reaction in the cascade (see Schemes 1 and 2). Indeed, had we kept the back reaction, with rate constant $k_{-S}$, in Equation (4), the denominator in Equations (9) and (10) would be replaced by $k_S S_0 + k_U U_0 + k_{-S}$. The phenomenological parameterization of Equation (12) would then become two-parameter, with the denominator in Equation (12) replaced by $x + ya + c$, where $c = k_{-S}/k_S S_{0,max}$. However, as mentioned earlier for MPh the specifics of the mechanism are not well studied, and the order of intake of the two substrates is not unique.[89,90] Therefore, other modifications of the simplest MM description would have to be considered, accompanied by an experimental study of the details of the MPh functioning under the considered system conditions, which is outside the scope of the present work.

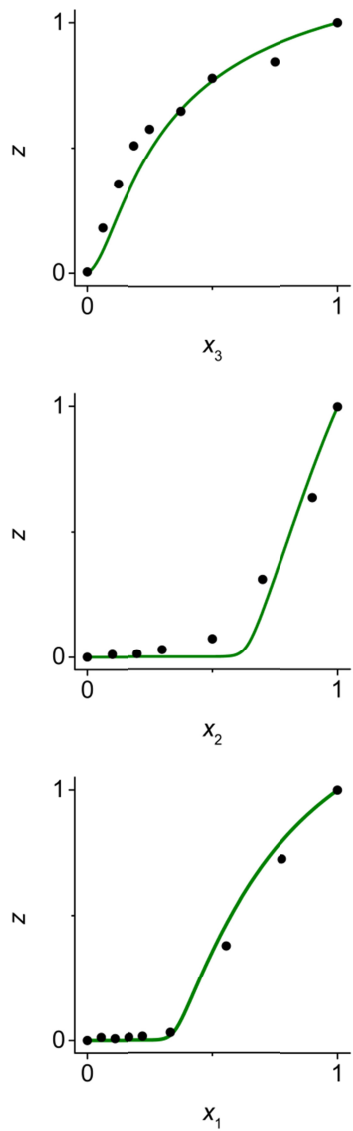

**Figure 4.** Experimental data (points) and theoretical curves for the variation of the absorbance as a function of different inputs, with all the quantities normalized to their logic ranges, similar to Figure 1, but with both the HK-catalyzed and NADH filter processes added, see Scheme 2. Note that the theoretical curves in all the panels are not fitted but rather calculated by using parameter estimates obtained earlier; see text for details.

**CONCLUSION**

At the level of individual network elements, in this work we derived a new single-parameter parameterization for two-input enzymatic AND gates without filtering, Equation (12), which captures several earlier noticed properties. We also considered a flexible approach to adding the filtering description by phenomenological closed-form expressions, involving separating out the signal that is filtered, as being processed via an additional identity gate. The latter can then be modified to introduce the filter-process parameters, exemplified by replacing the linear step in Equation (16) with Equation (23). The proposed phenomenological functions performed reasonably well in fitting some experimental data sets to determine the parameters in groups of one or two at a time, as well as in reproducing other data sets with these parameters, without any other adjustments.

The present study is the first attempt to parameterize networked processes functioning as a small enzymatic cascade with added filtering. We offer evidence that scaled (to reference ranges) "logic variables" for the inputs, output and some intermediate products can be useful in describing enzyme cascade behavior by identifying quantities that offer the most direct control of the network properties, and also allowing to approximately fit the system's responses with fewer adjustable parameters. While this approach is at best semi-quantitative and should be used with caution, we note that it is useful beyond the context of the "binary logic" network applications. The most obvious non-binary application could be to make some of the network responses as linear as possible for predefined ranges of inputs, which is of interest in certain sensor development situations.

**ACKNOWLEDGEMENTS**

Funding of our research by the NSF, via awards CCF-1015983 and CBET-1066397, is gratefully acknowledged.